\documentclass[12pt,showpacs,preprintnumbers,amsmath,amssymb,floatfix]{revtex4}
\usepackage[pdftex]{}
\usepackage{epsfig} 
\usepackage {array} 
\usepackage {amsmath} 
\usepackage {lscape}
\usepackage{graphicx}

\providecommand\bU{\boldsymbol{\rm U}}

\providecommand\rm{R_m}
\begin{document}
\preprint{}
\title{An optimal scale separation for a dynamo experiment}
\author{Franck Plunian}
\email{Franck.Plunian@hmg.inpg.fr}
\homepage{http://legi.hmg.inpg.fr/~plunian}
\affiliation{
Laboratoires des Ecoulements G\'{e}ophysiques et
Industriels,B.P. 53, 38041 Grenoble Cedex 9, France \\
}
\date{\today}
\begin{abstract}
Scale separation between the flow and the magnetic field is a common feature of natural dynamos. It has also been used in the Karlsruhe sodium experiment in which the scale of the magnetic field is roughly 7 times larger than the scale of the flow [R. Stieglitz and U. M\"uller, Phys. Fluids 13, 561 (2001)]. Recently, Fauve \& P\'etr\'elis [``Peyresq lectures on nonlinear phenomena", ed. J. Sepulchre, World Scientific, 1 (2003)] have shown that the power needed to reach the dynamo threshold in a dynamo experiment increases with the scale separation in the limit of large scale separation. With a more elaborate method based on subharmonic solutions [F. Plunian and K.-H. R\"adler, Geophys. Astrophys. Fluid Dynamics 96, 115 (2002)], we show, for the Roberts flow, the existence of an optimal scale separation for which this power is minimum. Previous results obtained by Tilgner [Phys. Lett. A 226, 75 (1997)] with a completely different numerical method are also reconsidered here. Again, we find an optimal scale separation 
in terms of minimum power for dynamo action. In addition we find that this scale separation compares very well with the one derived from the subharmonic solutions method.
\end{abstract}
\maketitle
We consider a dynamo experiment with a horizontal scale separation between the characteristic scale $l$ of the flow and the size $L$ of the container
as for example in the Karlsruhe experiment \cite{Stieglitz-01}.
In addition, the flow is assumed to have a geometry which can lead to the self-excitation of a magnetic field at the size of the container. 
For that we consider a Roberts \cite{Roberts-72} flow within a cubic box as in \cite{Plunian-02a}. In the $x$ and $y$ directions (where $x$, $y$ and $z$ are the cartesian coordinates), the size of a flow cell is $l \times l$ and the size of the box is $L \times L$. In the perpendicular direction $z$ the flow cells and the box have the common size $H$. Then the number of cells is $N_c = L^2 / l^2$.\\
Following \cite{Fauve-03}, we assume that the power $P$ is dissipated by turbulence, leading to $P=\rho L^2 H U^3 / l$ where $\rho$ is the density and $U$
the characteristic speed of the fluid. Defining the magnetic Reynolds number by $R_m = Ul/ \eta$, where $\eta$ is the magnetic diffusivity of the fluid, we find after some simple algebra that
\begin{equation}
	P = \rho \eta^3 \frac{H}{L^2}N_c^2 R_m^3.
	\label{Power}
\end{equation}

As a preliminary step, we assume that the first order smoothing approximation is valid (a sufficient condition being $R_m \le 1$). Then we have the relation $\eta b / l^2 \approx U B / l$ between the small scale $b$ and the large scale $B$ magnetic field intensities. Furthermore at the onset we have the following relation $\alpha K = \eta K^2$ derived from the mean part of the induction equation and where $\alpha$ corresponds to the anisotropic $\alpha$-effect produced by the Roberts flow, $K$ being the vertical wave number of the magnetic field. Here we take $K=1/H$, leading to $\alpha = \eta / H$.
Writing that the mean electromotive force $U b$ is equal to $\alpha B$ leads to the following relation
\begin{equation}
	U \sqrt{Hl} / \eta \approx 1.
	\label{SOCA}
\end{equation}
Then we can show that $\sqrt{\frac{H}{L}}R_m \approx N_c^{-1/4}$, leading to
\begin{equation}
	P \approx \frac{\rho \eta^3}{\sqrt{LH}}N_c^{5/4}.
	\label{SOCApower}
\end{equation}
From this simple estimate we conclude that the power consumption increases with the number of cells which is not in favor of scale separation. This was found previously by Fauve \& P\'etr\'elis \cite{Fauve-03} for a scale separation in the 3 cartesian directions (instead of only 2 in our case), leading to a different scaling $N_c^{5/6}$.
Both estimates are based on the first order smoothing approximation which has been proved to be too simplistic in the theoretical predictions of the Karlsruhe experiment. Therefore we reconsider this problem below in the light of the subharmonic solutions as studied in \cite{Plunian-02a}.\\ 

\begin{figure}[htbp]
\begin{tabular}{@{\hspace{1cm}}c@{\hspace{0cm}}c@{}}
\raisebox{7cm}{$P \cdot H $}&\epsfig{file=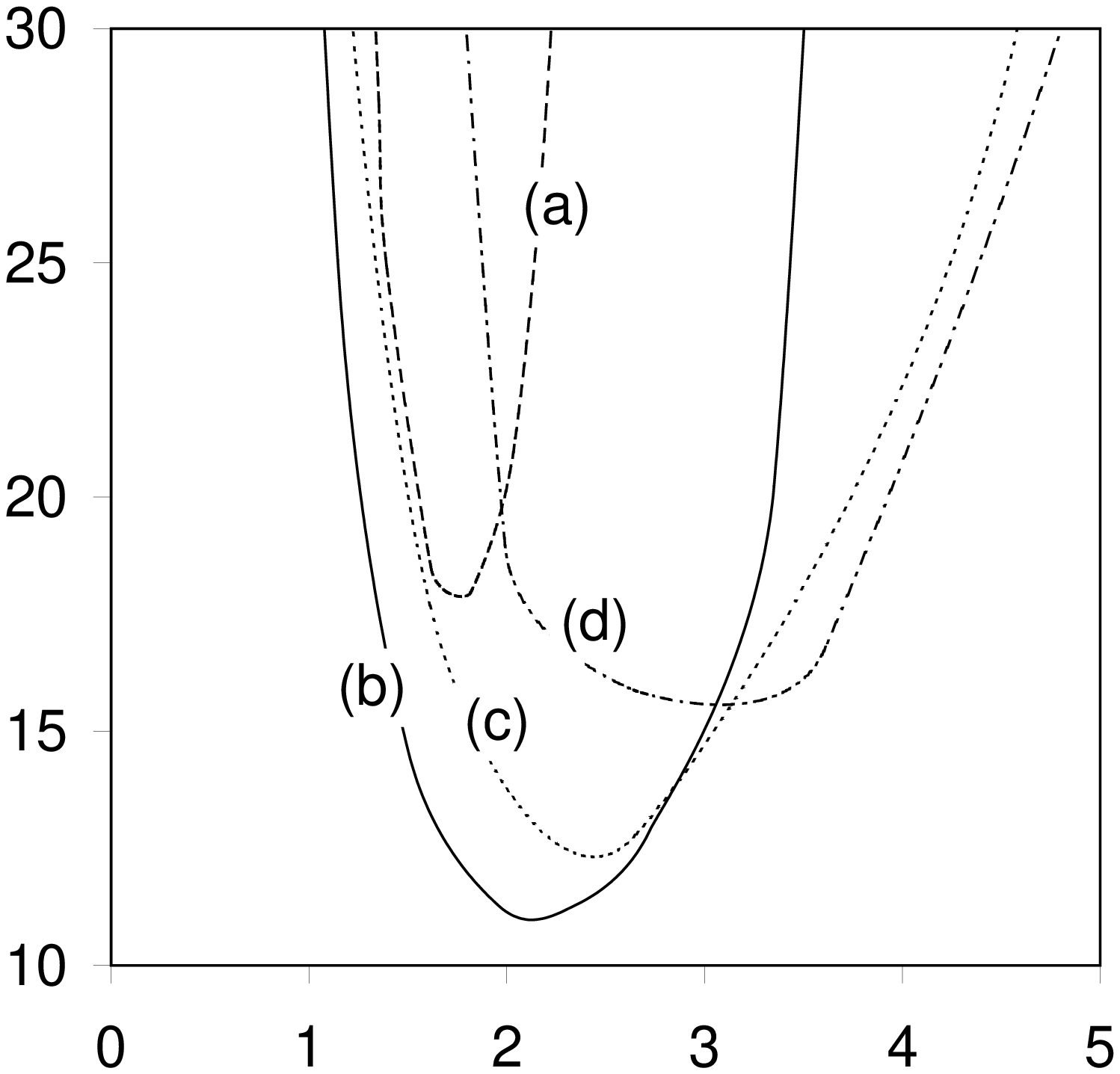,width=1\textwidth}\\*[-1.5cm]
&$L / H$
\end{tabular}
\caption{The consumption power $P \cdot H$ (in kW.m) versus $L/H$ for different numbers of cells (a) $N_c=8$, (b) $N_c=18$, (c) $N_c=32$, (d) $N_c=50$.}
\label{puissance}
\end{figure}

\begin{figure}[htbp]
\begin{tabular}{@{\hspace{1cm}}l@{\hspace{0cm}}c@{}}
\raisebox{7cm}{$P\cdot H$}&\epsfig{file=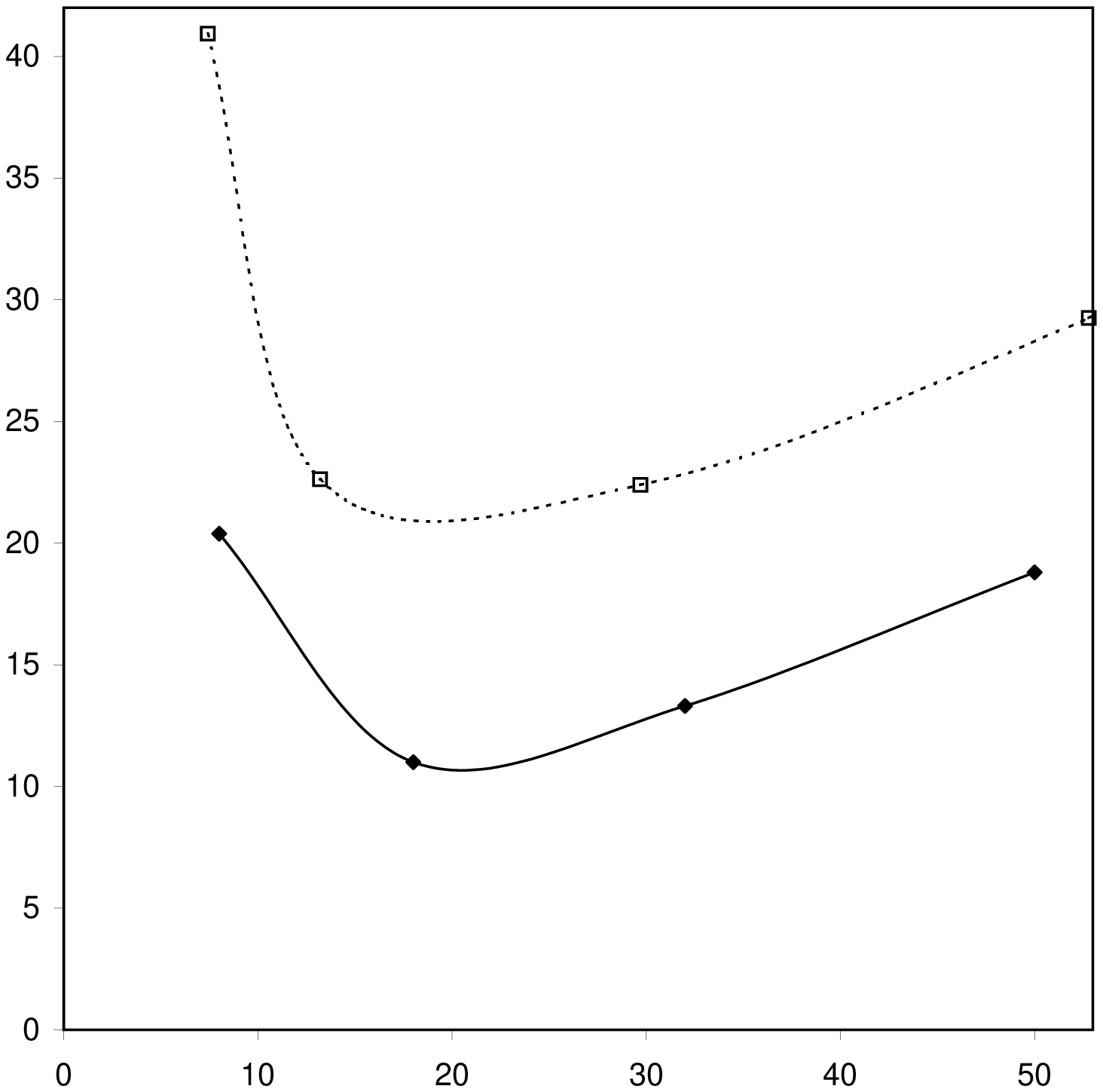,width=1\textwidth}\\*[-1.5cm]
&$N_c$
\end{tabular}
\caption{The consumption power $P \cdot H$ (in kW.m) versus the cells number $N_c$.
The full (dotted) curve is derived from \cite{Plunian-02a} (from \cite{Tilgner-97}) for $L/H=2$ ($R/H=1$).}
\label{puissance1}
\end{figure}

The original Roberts \cite{Roberts-72} flow is defined by
\begin{equation} 
\bU = U( \sin\frac{y}{L_U}, \sin\frac{x}{L_U}, \cos\frac{x}{L_U} 
-\cos\frac{y}{L_U} ) 
\label{robflow} 
\end{equation}
and the relations between the dimensions defined above and those defined in \cite{Plunian-02a} are $l=\pi \sqrt{2} L_U$, $L = \pi L_B$, $N=L_B/L_U$, $N_c = N^2 / 2$ and $R_m^* = R_m /(\pi \sqrt{2})$ where $R_m^*=U L_U / \eta$ is the magnetic Reynolds number defined in \cite{Plunian-02a}. For a given value of $N$, we look for the subharmonic solution embedded in the box of size $L\times L \times H$ and corresponding to the dimensionless wave numbers $f=1/N$ in the horizontal directions and $k=\frac{1}{N} \frac{L}{H}$ in the vertical direction. 
Then (\ref{Power}) writes in the form
\begin{equation}
	P \cdot H = \rho \eta^3 \frac{\pi^3}{\sqrt{2}} \left( \frac{N}{k} \right)^2 (R_m^*)^3 .
	\label{Power2}
\end{equation}
For a given value of $N$ the critical $R_m^*$ versus $k$ has been plotted in
Figure 4 of \cite{Plunian-02a}.
Then replacing $N$, $k$ and $R_m^*$ in (\ref{Power2}) we can calculate the corresponding power $P$ times $H$. For $\rho=10^3$ and $\eta=0.1$
we plot, in Figure, \ref{puissance} 
 $P.H$ (in kW.m) versus $k N = L / H$, for different values of $N_c$.
We find that the minimum value of $P\cdot H$ is obtained
for $L/H = 2.16$ and $N_c=18$ cells.
The case of the Karlsruhe experiment corresponds approximatively to $L/H = 2$ and $N_c=50$ for which we find $P\cdot H =19$ kW.m. For the same value of
$L/H$ but for $N_c=18$ cells, the power consumption is reduced roughly by a factor 2. 
In Figure \ref{puissance1}, $P\cdot H$ is plotted versus $N_c$ for $L/H=2$ (full curve). We see that there is indeed a minimum
around $N_c=20$ and that at large $N_c$, 
$P\cdot H$ increases with $N_c$ as predicted by (\ref{SOCApower}).
In a previous study \cite{Tilgner-97}, Tilgner calculated the critical magnetic Reynolds number for the Karlsruhe experiment geometry, varying the number of cells inside the device (Figure 4 of \cite{Tilgner-97}). The resolution was made with a completely different method than the one used in \cite{Plunian-02a} and it is then of interest to reconsider the results of \cite{Tilgner-97} in terms of power consumption and see how they compare to our results.
For that we need to make preliminary correspondance between our present notations and those used in \cite{Tilgner-97}.
In \cite{Tilgner-97} the flow container is a cylinder. then the consumption power, instead of (\ref{Power}), writes in the form
$P \cdot H = \frac{\rho \eta^3}{\pi} \left(\frac{H}{R}\right)^2 N_c^2 R_m^3$ where $R$ is the cylinder radius. In \cite{Tilgner-97} we have $l=\frac{8}{4.1} \frac{R}{N_{T97}}$ where we call here $N_{T97}$ the parameter $N$ of \cite{Tilgner-97}. This leads to a number of cells $N_c = \frac{\pi R^2}{l^2}=0.825 N_{T97}^2$. Furthermore the magnetic Reynolds number in \cite{Tilgner-97} is defined by $R_{mT97}=Ur_0/\eta$ where $r_0=\sqrt{1.25} R$ is the radius of the conducting sphere in which the cylinder is embedded. This leads to $R_m=\frac{1.745}{N_{T97}} R_{mT97}$.
Finally using the results from the Figure 4 of \cite{Tilgner-97}, the consumption power is plotted versus the cells number on
Figure \ref{puissance1} (dashed curve).
We find that there is again an optimal scale separation for which the dissipated power is minimum and again it corresponds to $N_c$ close to 20.
Furthermore the levels of power are of the same order of magnitude. We could not expect better agreement as the geometries and boundary conditions of \cite{Plunian-02a} and \cite{Tilgner-97} are really different.
Now considering the design of the Karlsruhe experiment, most of the dissipation power occurs in the pieces of pipes which redirect the flow into neighbouring cells at the end of each cell. The dissipation scaling in there is somewhat slower than $U^3$ but, most importantly, it is not proportional to the volume of the experiment. Therefore the scale separation of that experiment was guided by the characteristics of the available pumps \cite{Busse-96} in order to minimize the critical $R_m$, instead of minimizing the dissipated power, leading to $N_c=52$ (or alternatively to $N_{T97}=8$ corresponding to the minimum critical $R_m$ in \cite{Tilgner-97}). Therefore the criterion that we derived here is relevant for propeller driven experiments, but the requirements are more complicated (and also less universal) for pump driven experiments. 
\begin{acknowledgments}I am indebted to Jean-Fran\c{c}ois Pinton for having suggested this work, to Stephan Fauve for stimulating discussions at the Isaac Newton Institute during the Workshop on Magnetohydrodynamics of Stellar Interiors and to Andreas Tilgner 
for useful comments concerning the design of the Karlsruhe experiment.
\end{acknowledgments}

\end{document}